# Melting of iron close to Earth's inner core boundary conditions and beyond

**Authors:** M. Harmand[1,2]* and A. Ravasio[1], S. Mazevet[3,4], J. Bouchet[4], A. Denoeud[1], F. Dorchies[5], Y. Feng[6], C. Fourment[5], E. Galtier[6], J. Gaudin[5], F. Guyot[2], R. Kodama[7], M. Koenig[1], H.J. Lee[6], K. Miyanishi[7], G. Morard[2], R. Musella[3], B. Nagler[6], M. Nakatsutsumi[8], N. Ozaki[7], V. Recoules[4], S. Toleikis[9], T. Vinci[1], U. Zastrau[6,10], D. Zhu[6], A. Benuzzi-Mounaix[1,3]

**Affiliations:**

[1]LULI, Ecole Polytechnique, CNRS, CEA, UPMC, Palaiseau, France.

[2]IMPMC, CNRS, UPMC, MNHN, IRD, Paris, France.

[3]LUTH, Observatoire de Paris, CNRS, University Paris Diderot, Meudon, France.

[4]CEA-DAM-DIF, Arpajon, France.

[5]CELIA, Univ. Bordeaux, CEA, CNRS, Talence, France.

[6]LCLS, SLAC Stanford, USA.

[7] Graduate School of Engineering, Osaka University, Osaka, Japan.

[8]European XFEL, GmbH, Albert-Einstein-Ring 19, 22671 Hamburg, Germany

[9]FLASH, Desy, Hamburg, Germany.

[10]Institute for Optics and Quantum Electronics, Friedrich-Schiller-University of Jena, Germany.

*Correspondence to: marion.harmand@impmc.upmc.fr

**Abstract:**
Several important geophysical features such as heat flux at the Core-Mantle Boundary or geodynamo production are intimately related with the temperature profile in the Earth's core. However, measuring the melting curve of iron at conditions corresponding to the Earth inner core boundary under pressure of 330 GPa has eluded scientists for several decades. Significant discrepancies in previously reported iron melting temperatures at high pressure have called into question the validity of dynamic measurements. We report measurements made with a novel approach using X-ray absorption spectroscopy using an X-ray free electron laser source coupled to a laser shock experiment. We determine the state of iron along the shock Hugoniot up to 420 GPa (+/- 50) and 10800 K (+/- 1390) and find an upper boundary for the melting curve of iron by detecting solid iron at 130 GPa and molten at 260, 380 and 420 GPa along the shock Hugoniot. Our result establishes unambiguous agreement between dynamic measurement and recent extrapolations from static data thus resolving the long-standing controversy over the reliability of using dynamic compression to study the melting of iron at conditions close to the Earth's inner core boundary and beyond**.**

**Significant statement:**
The cores of telluric planets (Earth-type) are mostly composed of iron. Knowledge of molten or solid state at extreme pressure and temperature conditions of those metallic cores is necessary to understand temperatures profiles and magnetic field generation. The iron melting temperature is currently extrapolated at the Earth inner core boundary (330 GPa) from data obtained by static compression up to 150 GPa. Higher pressure measurements using shock compression were performed in previous works but highly questionable due to absence of direct detection of the molten state. Here, we couple laser shock with X-ray absorption spectroscopy to detect iron melting up to 420 GPa. We answer an important controversy, which affected shock measurements of the iron melting curve at around 260GPa.

When modeling the Earth's interior, essential features lie in our knowledge of iron and of iron alloys physical properties at extreme pressures and temperatures. While the density profile of the Earth's interior is now rather well constrained from seismic data, the temperature at the boundary between the solid inner core and liquid outer core (ICB, Inner Core Boundary), where the pressure is estimated to be of 330 GPa, remains up to now largely uncertain. It corresponds to the melting temperature of an iron alloy containing a small but yet unconstrained amount of impurities [1]. As a reference, the melting temperature of pure iron at ICB pressure condition is thus one of the most important parameters of Earth and planetary interiors physics. While first principles density functional theory simulations have provided a refined estimate of the melting temperature of iron of 5000 K at ICB conditions [2,3], the situation remains rather unclear mostly due to decades of difficulties in experimental confirmations of these predictions. The two main experimental approaches are the Diamond Anvil Cell (DAC), which provides static pressures, and the dynamical compression techniques. The disagreement between the static compression extrapolation of the melting temperature of iron at the ICB conditions with the dynamic measurements at conditions closer to the ICB led to important critics on the validity of the dynamic compression techniques [5,6,7]. These critics need to be addressed as this approach is so far the unique method that can potentially reach ICB conditions and beyond, as needed for the modeling of planets larger than Earth.

On the static side, a significant advance was recently accomplished by coupling laser-heated DAC compression to X-ray diffraction. This provided unambiguous measurements of the melting temperature of iron for pressures up to 150 GPa [4] and pointed out that chemical contaminations [5] and fast recrystallization [4] could have been at the origin of the contradicting measurements reported earlier. On the other hand, there have been doubts that melting of iron could be properly measured in dynamical experiments. Possible shock overheating [6], the apparition of transient phases before melting [7,8] and overall the difficulties in diagnosing unambiguously the solid - liquid transition within the short time scales involved in a dynamical process are open issues that steamed the iron controversy. These arguments were brought forward because melting measurements using laser-induced dynamic compression were inferred indirectly using optical diagnostics on the surface of the shock front [8-10] or sound velocity measurements [11-12]. There is thus an urgent need to infer directly structural information in dynamically compressed iron state. In that sense coupling ultrafast time resolved X-ray diagnostics with laser-induced dynamic compression techniques to detect the melting close to the ICB conditions and beyond would definitely settle these issues.

Coupling X-ray diagnostics to dynamical compression has been an important preoccupation for decades [13-17]. The X-ray pulse duration must be shorter than the processes involved and than the time over which uniform high pressures and temperatures are maintained (typically on the order of 100 ps in laser driven shock waves). The brightness and the accessible spectral range of the X-rays are also potential limitations that must be addressed in order to probe samples from a few to hundreds of μm thicknesses. X-ray Free Electron Laser (XFEL) sources perfectly fulfill such requirements by providing femtosecond (fs) pulses with unique peak brightness ($10^{12-13}$ photons/pulses) up to 10 keV photon energies, and are thus able to probe the short lifetime thermodynamic conditions occurring during shock compression. In this context, we used the recently commissioned MEC (Matter in Extreme Conditions) instrument at the Linac Coherent Light Source (LCLS) in Stanford, USA, to provide ultrafast high-brilliance X-ray pulses to probe iron compressed by laser shocks to pressures of up to 420 GPa for durations of several tens of picoseconds. Fig. 1 shows the experimental setup that we used to investigate the state of iron along the principal shock Hugoniot [18]. Two long laser pulses (532 nm, 3 ns, from 2 to 7 J per pulse) were focused down to a 200 μm spot size on iron targets leading to intensities of $10^{12-13}$ W/cm$^2$. The target geometry and optical diagnostics (VISAR for Velocity Interferometer System for Any Reflector and SOP for Streaked Optical Pyrometer) for extraction of thermodynamical conditions are shown in figure 1. Further details on the design, analysis and error bars calculations are explained in the Method Section and Supplementary Material.

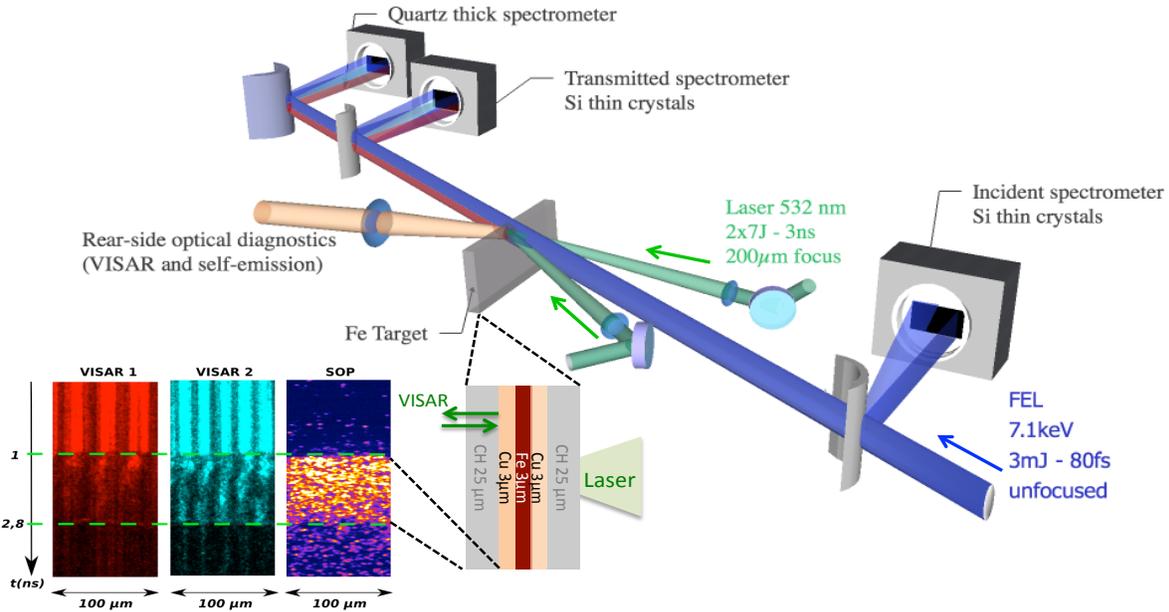

**Figure 1: Experimental setup at LCLS.** Two lasers are focused on the multi-layer iron sample while the unfocused LCLS beam probes the compressed sample within fs times scales. A first spectrometer measures the incident spectra while two different spectrometers are detecting the transmitted beam allowing XANES measurements with two different methods. A zoom of the target is displayed with images from VISAR and SOP measurements.

Following recent progresses using X-ray Absorption Near Edge Structure (XANES) spectroscopy in dynamical experiments for obtaining information at both the electronic and structural levels of shock compressed states [15,16,17], we used XANES spectroscopy to follow the evolution of the structural state of iron along the principal Hugoniot. Signatures in the XANES spectra were predicted using first principles simulations [20] and compared to previous measurements in liquid iron using synchrotron radiation coupled to static compression [19,21,22]. Compared to X-ray diffraction actively pursued at the moment in dynamical experiments, the X-ray absorption technique offers the advantage of requiring a lower photon number and thinner samples that are easier to compress homogeneously. It requires broad-band X-ray sources and has been coupled to laser based dynamic compression only recently [14-17]. These latter experiments were using ps secondary X-ray sources either limited to few keV when using M-band emission [23] or requiring ultra-high energy lasers (40 beams at the Omega facility [14]). More recently, our experimental team has demonstrated the possibility to perform XANES spectroscopy over few tens of eV around an absorption edge by using free electron laser radiation accumulated over few tens of single-shots only [24]. Here, we use simultaneously two different methods based on shot-to-shot measurement of the incident spectra (see Method Section). Due to the sharp spectral features of the XFEL source [24, 25], we accumulated several single-shots measurements to obtain a XANES spectrum extending 30 eV beyond the iron K-edge. Consequently, to obtain a reliable spectrum, we averaged over similar (P,T) conditions and extracted 5 distinct measurements along the Hugoniot up to 420 GPa and along the release adiabat at 12 GPa (see Supplementary Material). The high temporal resolution from the XFEL pulse duration brings confidence in probing homogeneous shock conditions.

Figure 2-(A) shows the complete set of measured XANES spectra obtained along the principal shock Hugoniot and one along the release adiabat. Here, we show the complete set of data for the membrane spectrometer only but the data sets of the two spectrometers are consistent and lead to the same conclusion (see Supplementary materials). At ambient conditions, the XANES spectrum is in very good agreement with standard measurements at synchrotrons. Figure 2-(B) shows the theoretical XANES spectra obtained at the corresponding conditions and considering each time either a solid Hexagonal Close Packed (HCP) or a liquid Fe phase. The calculated spectra were obtained by first performing molecular dynamics simulation based on the generalized gradient formulation of density functional theory for the given solid and liquid phases expected at a given condition, following the method of reference [20]. The XANES spectrum along the K–edge was calculated using linear response theory on various snapshots along that trajectory. We see a remarkable agreement with the experimental data for the variation of the XANES spectra in both density and temperature.

In figure 2-(A) and -(B), the distinct XANES spectra obtained for each phase allow a direct identification of the state of iron along the principal Hugoniot. For the 130 GPa (+/- 13) – 2780 K (+/- 230) (blue line), the XANES spectrum presents a strong structuration with a pre-peak at ~7.12 keV that was identified as an effect of the hybridization of the 3d-4p bands [26]. This feature is the signature of an HCP phase that has been well documented for instance at 20 GPa and 78 GPa [21,22] and predicted at higher pressures [20]. We also identified two XANES spectra, at 12 GPa (+/- 8) – 4180 K (+/- 350) and 420 GPa (+/- 50) – 10800 K (+/- 1390) (purple and red lines) that have distinct spectral features. The disappearance of the edge structure mainly around 7.12 keV as well as a change of the slope are systematically observed even far above the predicted high-pressure melting curve, up to 420 GPa, and at 12 GPA – 4180 K where iron is known to be liquid. These observations are very consistent with those of previous DAC experiments at lower pressure [22]. Experimental XANES spectra from melted iron and XANES calculations of solid HCP and melt at the same thermodynamical conditions allow us to understand the disappearance of the 7.12 keV pre-peak and the change of the edge slope as a clear signature that the melting line was crossed.

In figure 2-(C) and -(D), we compare the XANES spectra obtained for the body cubic centered (BCC) phase at ambient condition with the 130 GPa – 2780 K and the 260 GPa – 5680 K conditions. The two XANES spectrometers are in good agreement both showing the strong pre-peak feature at 7.12 keV as well as its disappearance at 260 GPa. Figure 2-(E) shows calculated XANES spectra for the same conditions with the same behavior being observed. In both experimental and simulated XANES spectra, the structure change in the edge affords a clear signature of the melting and allows us to unambiguously identify the 260 GPa (+/- 29) – 5680 K (+/-700) and 380 GPa (+/-30) – 8120 K (+/-1400) conditions as melted iron. The disappearance of the 7.12 keV pre-peak is the signature of the change of the electronic density of state [26], which is highly sensitive to the solid – liquid transition of iron. This criterion is expected to be more sensitive to the melting transition than changes in the optical reflectivity, which can sometimes be due to recrystallization in the solid state or than the loss of structuration in X-ray absorption spectra, which can be inadequate because of the persistence of a short-range order in the liquid phase [27].

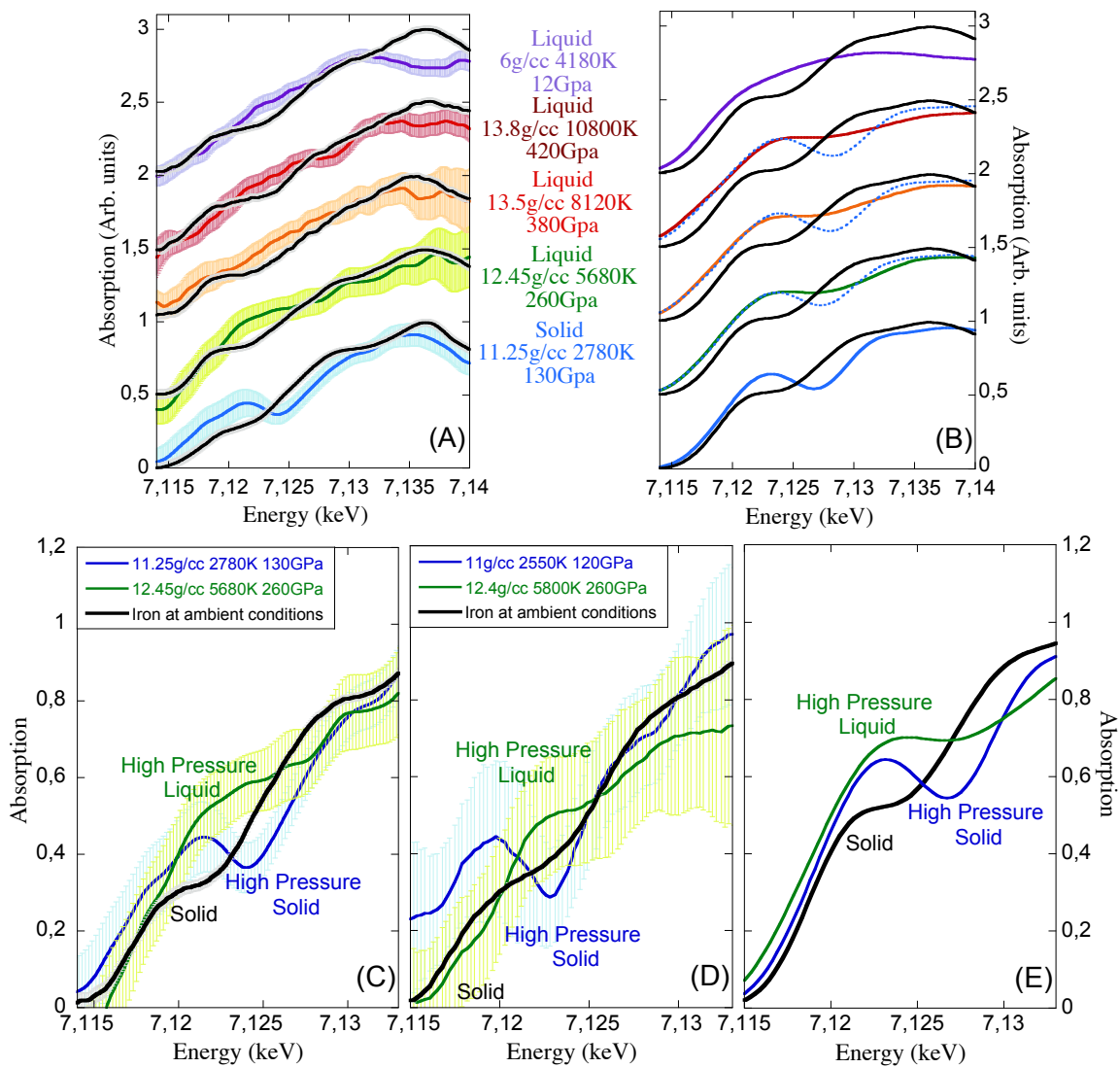

**Figure 2: X-ray absorption spectroscopy (A)** XANES spectra with error bars (see Suppl. Material) measured at various conditions compared to spectra at ambient conditions and for the corresponding samples (black). **(B)** Corresponding *ab-initio* calculations for liquid and hcp (dotted lines) phases compared to ambient conditions (black). **(C)** Experimental XANES spectra at ambient conditions (black), compressed solid iron at 130 GPa (blue) and close to the melting curve at 260 GPa (green), obtained with the membrane spectrometers diagnostic. **(D)** XANES spectra obtained with the quartz spectrometer at similar conditions (not averaged on the same shots) **(E)** Corresponding *ab-initio* calculations.

In figure 3, we show the iron phase diagram where we report the latest experimental and theoretical data as well as the results of our study. These results indicate a melting temperature below 5680 K at 260 GPa (upper limit of 5915 K deduced from the error bars) which is consistent with the extrapolation from recent DAC results [4] and with *ab initio* simulations [2,3] but significantly lower than estimates from previous dynamic measurement using optical diagnostics [9,10,29], which gave a Hugoniot melting temperature of 6700 K at 243 GPa [29]. Our measurements are consistent with previous data obtained from sound velocity measurements in shock experiments [11,12]. They thus clarify the results from previous dynamic compression experiments obtained in this region of the phase diagram by providing the first detection of a melt at these thermodynamic conditions, thus leaving no room for speculation regarding the underlying iron phase state under shock compression. We point out that experimental results on melting using an X-ray diagnostic probing bulk sample are less sensitive to shock overheating than optical surface diagnostics that probe only the shock front. Indeed, shock overheating occurs when the temperature rises faster than the rate of atom rearrangement required for melting [6]. While optical diagnostics may probe time scales corresponding to the rise of temperature in the shock front and therefore are more sensitive to kinetics, here we probe the bulk of the material a few hundreds of picoseconds after the shock breakout in Cu. In our case, shock overheating, if present, is a second order effect when using a bulk probe such as X-rays, while it is critical when using optical diagnostics [9,10,29].

While remaining along the principal Hugoniot, our results significantly extend the knowledge of the high pressure melting curve of iron for conditions close to the ICB conditions at 330 GPa and well beyond the most recent diamond anvil cell experiments in which melting was directly measured up to 150 GPa [4]. These results resolve the long-standing controversy on the validity of the dynamic compression by observing the melting using X-ray absorption spectroscopy of bulk iron. They allow to invalidate previous extrapolations of the melting curve at the ICB conditions from Yoo *et al.* and Williams *et al.* [9,12] and reinforce recent extrapolations from Anzellini et al. suggesting melting temperature of 6200 K at 330 GPa [4]. Moreover, by directly detecting the molten states of iron at 380 GPa – 8100 K and 420 GPa – 10800 K, the present work brings further constrains on the phase domain of liquid iron at extreme pressures. It opens perspectives for systematic and extensive studies of planetary systems at extreme high pressures, so far only achievable with dynamic compression technics.

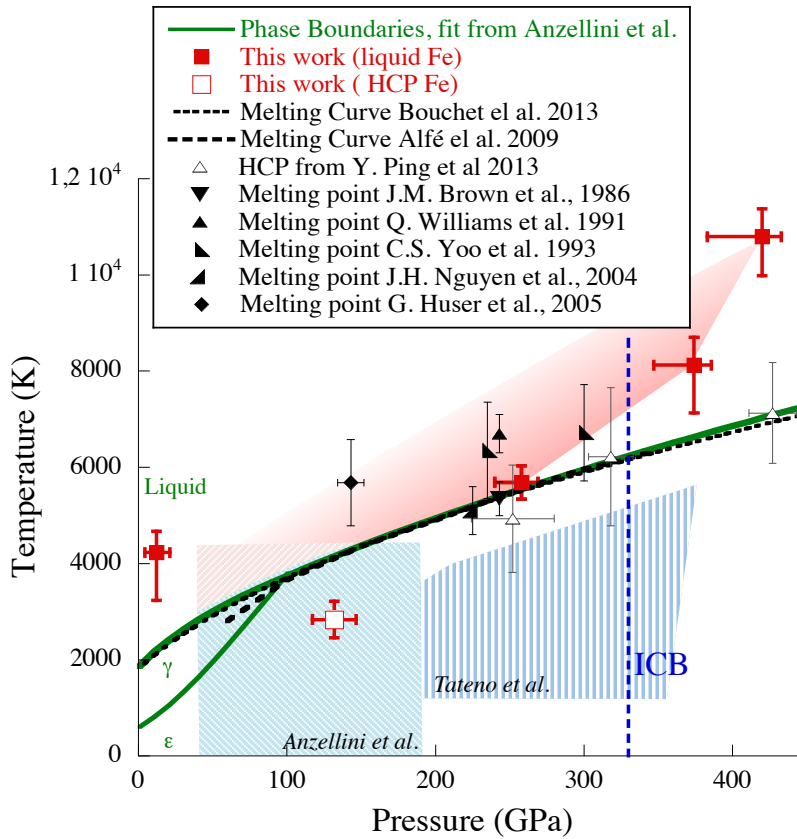

**Figure 3: Iron phase diagram:** Our results (red squares) are compared with previous results obtained in the literature from static compression [light blue], shock experiment [triangles], and ab-initio simulations [dashed lines]. The pink area shows an upper constrain line on the melting curve of iron from our measurements and previous diamond anvil cell extrapolations (green line) [4]. The Inner Core Boundary pressure of 330 GPa is shown with a dashed blue line.

**Method:**

**Laser shock compression:** The targets consisted of multilayer samples made of CH/Cu/Fe/Cu/CH. With this particular design, the two Cu layers keep the iron sample in a strong compression as long and uniform as possible. A first CH layer was chosen to optimize the ablation and to minimize x-ray radiation from the laser-produced coronal plasma while a rear side CH layer was used to infer the thermodynamic conditions of shocked state of copper and then iron using optical diagnostics. Fig. 1 shows an example of the Velocity Interferometer System for Any Reflector (VISAR) traces and of the time-resolved emission diagnostic signal obtained with a Streaked Optical Pyrometer (SOP), showing a very good spatial uniformity of the shock. From both the emission and the VISAR signals, we were able to measure, with a good precision, the mean shock velocity in the rear side CH layer using the shock transit time and the initial thickness. In addition and for high laser intensity, VISAR gave also the time-resolved shock velocity in CH. These measurements were coupled to hydrodynamic simulations using validated equation of states to infer the thermodynamical state of the iron sample at the probed time (see Suppl. Material). All these coupled measurements allowed us to extract reliable data for each laser shot. We estimated single-shot error bars by taking into account the timing uncertainty between the laser and the XFEL probe, estimated to be about 30 ps, the accuracy in the laser intensity, of 15%, uncertainty in the CH layer thickness of <10% and the choice of equation of state (see more details in the Suppl. Material). By probing the shocked iron at different times between the optical pump laser and the XFEL and by using different laser intensities, we scanned the phase diagram of iron over a large domain of pressure and temperature. The accumulation of statistics at specific conditions allowed us to significantly reduce errors which are usually important in dynamic compression experiments (see Suppl. Material).

**XANES measurements:** We used two simultaneous methods to measure the XANES spectra. The first one use bent crystals with the unfocused ~80 fs ~3 mJ 7.1 keV XFEL so that spectrometers are angularly dispersive along the horizontal direction but preserve the spatial information in the vertical direction. Coupled with a half-target design in the first method, only part of the beam goes through the sample while the remaining part of the beam is used for obtaining a reference spectrum. The extraction procedure associated with that method is further detailed in [24]. The second method consists in using thin identical Si crystals [25] placed symmetrically before and after the sample to directly measure the incident and transmitted spectra.

**Acknowledgments:** We acknowledge R. Torchio, P. Mounaix, D. Antonangeli and D. Cabaret for helpful discussions. This work is supported by the French Agence Nationale de la Recherche (ANR) with the ANR IRONFEL and Planetlab. This work was performed at the Matter at Extreme Conditions (MEC) instrument of LCLS, supported by the DOE Office of Science, Fusion Energy Science under contract No. SF00515. This work was also supported by LCLS, a National User Facility operated by Stanford University on behalf of the U.S. Department of Energy, Office of Basic Energy Sciences. U. Zastrau is grateful to the Volkswagen Foundation.